\documentstyle[twocolumn,prl,aps,epsf]{revtex}

\begin{document}
\draft

\title{
Truncation of a 2-dimensional Fermi surface\\
 due to quasiparticle gap formation
 at the saddle points
}
\author{Nobuo Furukawa\cite{PermAddress} and T. M. Rice}

\address{
  Institute for Theoretical Physics, ETH-H\"onggerberg,
  CH-8093 Zurich, Switzerland
}

\author{Manfred Salmhofer}
\address{
Mathematik, ETH Zentrum, CH-8092
Z{\"u}rich, Switzerland
}

\date{June 12, 1998}

\maketitle

\begin{abstract}
We study a two-dimensional Fermi liquid with a Fermi surface
containing the saddle points $(\pi,0)$ and $(0,\pi)$.
Including Cooper and Peierls channel contributions leads to a one-loop
renormalization group flow to strong coupling for short range
repulsive interactions. In a certain parameter range the characteristics 
of the fixed point, opening of a spin and charge gap and dominant
pairing correlations are similar to those of a 2-leg ladder at
half-filling. An increase of the electron density we
argue leads to a truncation of the Fermi surface 
with only 4 disconnected arcs remaining.

\end{abstract}

%


The origin of the instability of the Landau-Fermi liquid state
as the electron density is increased in overdoped cuprates
is one of the most interesting open questions in the field.
Recently we proposed that the origin lies in a flow of Umklapp scattering
to strong coupling \cite{Furukawa98}.
The simpler case with the Fermi surface (FS) extrema at
$(\pm\pi/2,\pm\pi/2)$ was considered and not the realistic case for 
hole-doped cuprates where the leading contribution from Umklapp processes
comes from scattering at the saddle points $(\pi,0)$ and $(0,\pi)$.
In this letter we report a one-loop renormalization group (RG)
calculation  for the realistic case including contributions from both
Cooper and Peierls channels. Reasonable conditions can lead to a
strong coupling fixed point 
whose characteristics are similar to those
of half-filled 2-leg ladders. There strong coupling Umklapp processes lead to
spin and charge gaps but only short range spin correlations.
A particularly interesting and novel feature is that although the
strongest divergence is in the d-wave pairing channel, the charge
gap causes insulating  not superconducting behavior.

There have been a number of previous RG investigations for
 a FS with saddle points.
Schulz\cite{Schulz87} 
and Dzyaloshinskii\cite{Dzyaloshinskii87}
considered the special case 
with only nearest neighbor (n.n.) hopping so that the
saddle points coincides with a square FS
and perfect nesting exactly at half-filling,
leading to a 
fixed point with long range antiferromagnetic (AF) order.
Lederer {\em et al.} \cite{Lederer87} and 
Dzyaloshinskii \cite{Dzyaloshinskii96} also considered the same model as we do.
There are two fixed points, one  at a strong coupling
fixed point with d-wave pairing found by Lederer {\em et al.} \cite{Lederer87},
and  a weak coupling examined by Dzyaloshinskii \cite{Dzyaloshinskii96}.
A Hubbard parametrization  of the repulsive interactions ($U$) and moderate
interaction strength suffices to stabilize the strong coupling
fixed point. The new feature  we  wish to stress is that there can
be both spin and charge gaps. The FS is then truncated through
the formation of an insulating spin liquid (ISL) with 
resonance valence bond (RVB) character. 
We propose that as the hole doping decreases these gaps spread out
from the saddle points so the FS consists of 
a set of arcs, which
progressively shrink as the hole doping decreases.

\begin{figure}
\epsfxsize=7.5cm
\hfil\epsfbox{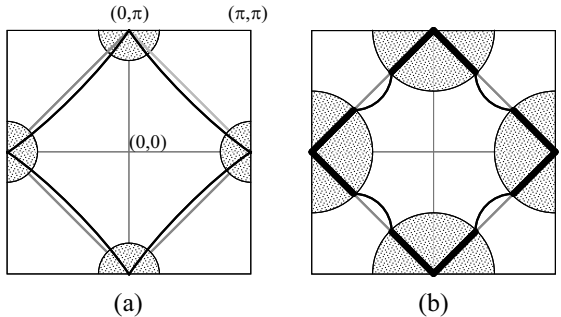}\hfil
\caption{Fermi surface (FS). (a) Two patches of the FS at the saddle points.
(b) Truncated FS as electron density is increased.}
\label{FigFS}
\end{figure}

We start with a 2-dimensional FS 
touching the saddle points $(\pi,0)$ and $(0,\pi)$.
Such a FS is realized in the case
of the dispersion relation 
$\varepsilon({\bf k}) = - 2 t (\cos k_x + \cos k_y) -
4 t' \cos k_x \cos k_y$ with $t >0$ ($t'<0$) as n.n. (n.n.n.)
hoppings.
Throughout this letter, we assume $t'/t$ small but nonzero
so that we are close to half-filling.
Due to the van Hove singularity, the leading singularity
arises from electron states in the vicinity of the saddle points.
We consider two FS patches  at the saddle points
and examine the coupling between them using one-loop RG equations,
as illustrated in Fig.~\ref{FigFS}a.
$k_{\rm c}$ is the radius
of the patches.

The susceptibility for the Cooper channel at $q=0$ has a
log-square behavior of the form
\begin{equation}
  \chi^{\rm pp}_0(\omega) = - h
       \ln({\omega}/{E_0})
       \ln({\omega}/{2tk_{\rm c}{}^2}).
\end{equation}
Here, the sum over $k$ is restricted to the patches.
$E_0$ is the cutoff energy and $h= {(8\pi^2 t)}^{-1}$ for $|t'/t| \ll 1$.
The Peierls channel
 at $Q=(\pi,\pi)$ diverges as
\begin{equation}
 \chi^{\rm ph}_Q(\omega) = 
  \left\{
   \begin{array}{ll}
     h \ln(\omega/E_0) \ln(\omega/2tk_{\rm c}{}^2) 
          & \quad \omega \gg |t'| \\
    2h \ln|t'/t| \ln(\omega/E_0 )&  \quad  \omega \ll |t'|
   \end{array}
   \right.  .
\end{equation}
The susceptibilities for the 
Peierls  channel  at $q=0$ ($\chi^{\rm ph}_0$) and
the Cooper channel at $q=Q$ ($\chi^{\rm pp}_Q$) also
diverges
\begin{equation}
 \chi^{\rm ph}_0 \sim -\chi^{\rm pp}_Q \sim 
  2h   \ln(E_0/\omega),
\end{equation}
but the coefficients of  $\ln(\omega)$ 
are smaller than that of $\chi^{\rm ph}_Q$.

\begin{figure}
\epsfxsize=7.5cm
\hfil\epsfbox{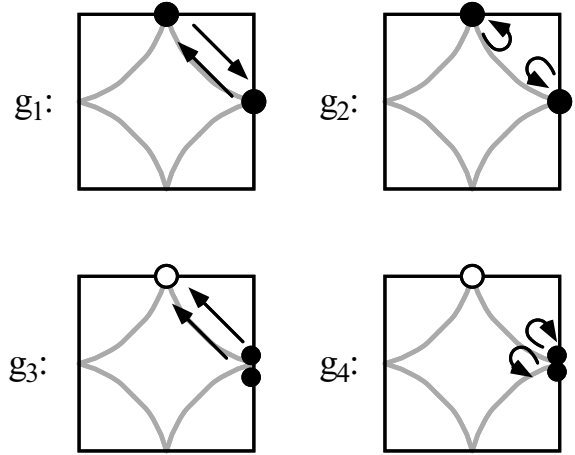}\hfil
\caption{The definitions of vertices for the 2-patch model.}
\label{FigVertex}
\end{figure}

In Fig.~\ref{FigVertex} we define the interaction vertices
$g_i$ ($i=1\sim4$).
Normal and Umklapp processes are indistinguishable
since the patches are at the zone edge.
We use a Wilson RG flow, parametrized by a decreasing energy
scale, in which all degrees of freedom above that energy scale
are integrated out. Wilson's effective action at scale $E_0$ 
has the dual interpretation
that (a) it generates the interaction vertices, and thus an effective
Hamiltonian, for the particles with energy below $E_0$
and (b) these vertices are also the connected correlation functions
with the infrared cutoff $E_0$. 
We consider only the four-point function and include
only the one-loop terms. The one-loop RG was justified as the
leading behavior at low energies and weak coupling for a class of FS in 
ref.~\onlinecite{Feldman96},
which includes those with nonzero curvature ($t' \ne 0$ in our case). 
It leads to the  flow equations
(see also Lederer {\em et al.} \cite{Lederer87})
\begin{eqnarray}
 \dot{g}_1 &=& 2 d_1 g_1 ( g_2 - g_1) 
             + 2 d_2  g_1 g_4
	     - 2 d_3 g_1 g_2
  \label{EqRG1}
  ,\\
 \dot{g}_2 &=& d_1 ( g_2{}^2 + g_3{}^2) 
             + 2 d_2 ( g_1 - g_2 ) g_4 
	     - d_3 (g_1{}^2 + g_2{}^2)
  \label{EqRG2}
  ,\\
 \dot{g}_3 &=& -2 g_3 g_4 
              + 2 d_1 g_3 (2 g_2 - g_1 )
  \label{EqRG3}
  ,\\
 \dot{g}_4 &=& -(g_3{}^2 + g_4{}^2)
             + d_2 ( g_1{}^2 + 2 g_1 g_2 - 2 g_2{}^2 + g_4{}^2)
  \label{EqRG4}.
\end{eqnarray}
Here we introduced the normalization
$g_i \to h g_i$ to give dimensionless couplings,
 and $\dot{g}_i \equiv ({\rm d}g_i)/({\rm d}y)$ where
$y \equiv \ln^2(\omega/E_0) \propto  \chi^{\rm pp}_0(\omega)$.
 We define 
 functions which describe the relative weight of $q=0$ Cooper channel 
contribution and those of other channels
\begin{eqnarray}
  d_1(y) &=& {{\rm d}\chi^{\rm ph}_Q}/{{\rm d} y},
   \\
  d_2(y)&=& {{\rm d} \chi^{\rm ph}_{0}}/{{\rm d}y},
   \\
  d_3(y) &=& -{{\rm d} \chi^{\rm pp}_Q}/{{\rm d}y}.
\end{eqnarray}
Their asymptotic forms are $d_1(y)\to 1$ at $y\approx 1$
and $d_1(y) \sim \ln|t/t'|/\sqrt{y}$ as $y \to \infty$,
while $d_2(y)\sim d_3(y)\sim 1/\sqrt{y}$ throughout the
region of interest.

The case $d_1 = 1$ and $d_2,d_3 \ll d_1$ 
was studied by Schulz\cite{Schulz87}, Dzyaloshinskii \cite{Dzyaloshinskii87}
and Lederer {\em et al.}\cite{Lederer87} which
arises at $t'=0$ as well as in a sufficiently large $U$ region
where $t'$ is irrelevant.
SDW susceptibility 
has the same exponent as d-wave pairing
but is dominant due  to the next leading divergent terms.
The fixed point is understood as a Mott insulator with long range AF order.
The limit $d_1=d_2=d_3=0$ was treated by 
Dzyaloshinskii \cite{Dzyaloshinskii96}.
In this case (\ref{EqRG3}) and (\ref{EqRG4}) combine to give
$
 \dot g_{-} = - g_{-}{}^2
$
with $g_{-} = g_4 - g_3$.
Dzyaloshinskii considered the  weak-coupling fixed point
$g_{-}\to 0$ which arises when
$g_{-} \ge 0$, and discussed the resulting Tomonaga-Luttinger liquid behavior.

\begin{figure}
\epsfxsize=7.5cm
\hfil\epsfbox{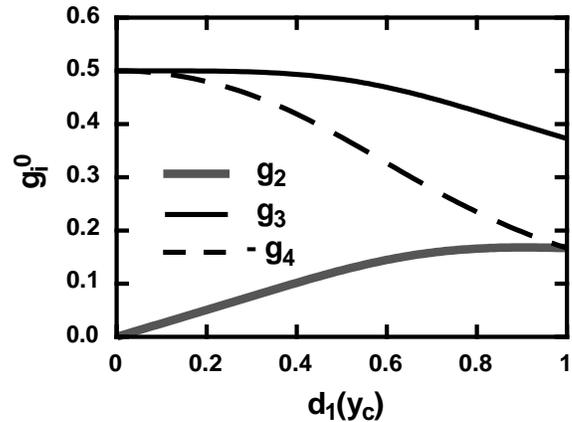}\hfil
\caption{The fixed point values for $g_i^0$.}

\label{FigGi0}
\end{figure}

In this letter we examine the RG equations with
$0<d_1(y)<1$ which enables us to consider 
nonzero values of the ratios $t'/t$ and $U/t$.
Since $d_2,d_3 \ll d_1$,
we neglect $d_2$ and $d_3$ in RG equations for simplicity.
Note the terms involving $d_1$ act to enhance the
basin of attraction for the strong coupling fixed point, $g_{-}\to -\infty$
\cite{Lederer87}.
The one-loop RG equations are solved numerically.
 Starting from
a Hubbard-model initial value $g_i= U$ ($i=1\sim4$), 
the vertices flow to strong coupling fixed points with
$g_2\to+\infty$, $g_3\to+\infty$ and $g_4\to-\infty$, with the
asymptotic form
\begin{equation}
 g_i(y) = {g_i^0}/({y_{\rm c}-y}).
 \label{EqGi0}
\end{equation} 
Here
$y_{\rm c} \sim t/U$ is the critical point of one-loop RG equations.
The divergence of $g_1(y)$  with respect to $y_{\rm c}-y$
is only logarithmic.
To analyze this fixed point more precisely, we 
substitute the asymptotic form  (\ref{EqGi0}) into
eqs.~(\ref{EqRG1})-(\ref{EqRG4}) and obtain polynomial equations
\begin{eqnarray}
 {g}_1^0 &=&
  2 d_1(y_{\rm c})\cdot g_1^0 ( g_2^0 - g_1^0)
  \label{EqFP1}
  ,\\
 {g}_2^0 &=& 
    d_1(y_{\rm c}) \cdot ( (g_2^0)^2 + (g_3^0)^2)
  \label{EqFP2}
  ,\\
 {g}_3^0 &=& -2 g_3^0 g_4^0 
    + 2 d_1(y_{\rm c}) \cdot g_3^0 (2 g_2^0 - g_1^0 )
  \label{EqFP3}
  ,\\
 {g}_4^0 &=& 
 -((g_3^0)^2 + (g_4^0)^2)
  \label{EqFP4}
\end{eqnarray}
Fig.~\ref{FigGi0} shows
the solution of these equations $g_i^0$ 
for the  initial values $g_i=U$.
The coefficients $g_i^0$ are determined as a function of
$d_1(y_{\rm c})\sim \sqrt{U/t}  \ln|t/t'|$, {i.e.}
the critical behavior of the fixed point is 
 a function of $U$.

Although one cannot solve for the strong coupling fixed point using 
only one-loop RG equations, a qualitative description
comes from the susceptibilities.
Using these coefficients $g_i^0$, 
exponents for various susceptibilities are calculated as follows.
The one-loop RG eqn. for  the d-wave pairing  is
\begin{equation}
 \dot{\bar \chi}_{\rm dP} = 2(g_4-g_3),
\end{equation}
where $\bar \chi_{\rm dP} = (\partial \chi_{\rm dP}/\partial \omega)
/(\partial \chi^{\rm pp}_0/\partial \omega)$. From eq.~(\ref{EqGi0}), we 
obtain a divergence  $\chi_{\rm dP} \propto (y_{\rm c}-y)^{\alpha}$
with exponent $\alpha=\alpha_{\rm dP}= 2(g_4^0 - g_3^0)$.
Similarly, exponents for s-wave pairing,
charge (CDW), spin (SDW) density waves
as well as uniform spin, charge compressibilities and
finite momentum $\pi$-pairing
are given by
\begin{eqnarray}
  \alpha_{\rm sP} &=& 2(g_3^0+g_4^0),\\
  \alpha_{\rm CDW} &=& (2 g_1^0-g_2^0+g_3^0)\, d_1(y_{\rm c}),\\
  \alpha_{\rm SDW} &=& -2(g_2^0+g_3^0)\, d_1(y_{\rm c}),\\
  \alpha_{\rm s} &=& -2(g_1^0 + g_4^0)\, d_2(y_{\rm c}),\\
  \alpha_{\kappa} &=& ( -g_1^0+2 g_2^0+g_4^0)\, d_2(y_{\rm c}),\\
  \alpha_{\pi} &=& 2(-g_1^0+ g_2^0) \, d_3(y_{\rm c}),
\end{eqnarray}
respectively. 
For weak coupling, we have
 $d_2(y_{\rm c}) \sim d_3(y_{\rm c}) \sim \sqrt{U/t}$.
Uniform susceptibilities are calculated in the limit
$\omega,q \to 0$ with $q/\omega$ held fixed.

\begin{figure}
\epsfxsize=7.5cm
\hfil\epsfbox{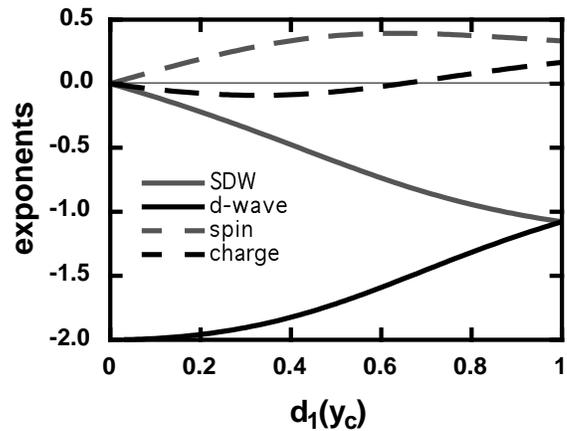}\hfil
\caption{Exponents for various susceptibilities. For uniform
spin and charge susceptibilities, exponents are scaled by $d_2/d_1$}
\label{FigExpo}
\end{figure}

In Fig.~\ref{FigExpo} we show the exponents for
d-wave pairing, SDW, uniform spin and charge compressibility.
Comparison of the values of the exponents shows us that 
the most divergent susceptibility is  d-wave pairing
throughout the parameter region of $0<d_1(y_{\rm c})<1$ \cite{Lederer87}.
The SDW susceptibility shows a weaker divergence and 
the exponent vanishes in the limit $d_1(y_{\rm c})\to 0$ or $U/t\to 0$.
The exponents for uniform spin as well as
s-wave pairing, CDW and $\pi$-pairing  are always positive,
i.e. these susceptibilities are suppressed
at low frequency.

The exponent for the charge compressibility changes 
sign at $d_1(y_{\rm c}) \sim 0.6$. Namely there exists a critical
interaction strength $U_{\rm c}$ such that for $U>U_{\rm c}$
the charge compressibility is suppressed to zero.
The critical value $U_{\rm c}$ is determined by $t'$ in the
form $U_{\rm c}/t \propto \ln^{-2}|t/t'|$.
This implies a
transition from a superconducting phase at $U<U_{\rm c}$
with its origin in enhanced Cooper pairing
due to the van Hove singularity,
to a charge-gapped phase at $U>U_{\rm c}$ which can be regarded as a
precursor of the Mott transition.
The fixed point at $U>U_{\rm c}$ resembles that of the
half-filled 2-leg ladder which has spin and charge gaps
but the most divergent susceptibility is d-wave pairing.
This fixed point (C0S0 in the
Balents-Fisher\cite{Balents96} notation) 
is well understood as an ISL of short
range RVB form. The close similarity between the fixed points
leads us to assign them to the same universality class.

The fixed point for $0<d_1 <1$
with  d-wave pairing as the leading divergence
differs from the case $d_1 \equiv 1$ where
SDW correlation is dominant when the  next leading term is included
\cite{Schulz87}, and
from the weak-coupling fixed point for $d_1\equiv 0$  \cite{Dzyaloshinskii96}.
The contribution of the particle-particle channel of the part 
of the Fermi surface away from the saddle points was not discussed here
for brevity; one can show that it gives a nonnegative contribution to
$\dot g_{-}$ and thus even enhances the asymmetry that drives
the flow to strong coupling $g_{-}\to-\infty$. 
%
%
%
The present result for the saddle point model with $t'<0$
is in good accordance with the case $t'>0$ (or, $t'<0$ with
electron doping)
previously studied by the authors \cite{Furukawa98}.
There, 4 patches at the zone diagonals were considered.
For interaction $U$ larger than a critical value determined
by the infrared cutoff due to finite curvature,
the charge compressibility renormalizes to zero
due to the Umklapp scattering.

Next we consider increasing the electron density. One possibility
is to follow the non-interacting FS which expands beyond
the saddle points.
 But the flow to strong coupling and the opening of a charge 
and spin gap leads us to consider a second possibility, namely
that the FS is pinned by Umklapp processes and does not expand beyond the
saddle points.
This proposal was put forward in Ref.~\onlinecite{Furukawa98}
after an examination of 8 FS-patches 
located on the Umklapp surface (US) which is defined by lines
joining saddle points. The leading contribution from Umklapp
processes comes from scattering between points on this US.
Support for this proposal comes from the lightly doped
3-leg ladder\cite{Rice97,White98}
 where in strong coupling a C1S1 phase occurs with an
ISL with exactly half-filling in the even parity channels and an
open FS only in the odd parity channel. This contrasts with the one-loop
RG results which gives a C2S1 phase\cite{Balents96}
 with holes immediately entering
both odd and even parity channels. Our proposal, sketched in
Fig.~\ref{FigFS}b, is based on a lateral spread of the
spin and charge gaps along the US leading to  4 open
FS segments consisting of arcs centered at the points
$(\pm\pi/2,\pm\pi/2)$.
Such behavior can be viewed as a sort of phase separation in $\vec k$-space
in that in some directions an ISL forms but others remain metallic.
 Note the area enclosed by the surface
defined by the US and these 4 arcs contains the full
electron density, consistent with a generalized form of 
Luttinger's Theorem.

Note since the ISL is not characterized by any simple broken
symmetry or order parameter, the resulting state cannot be
described by a simple mean field or Hartree-Fock factorization.
Our proposal of a FS consisting of 4 disconnected arcs has strong
parallels to recent gauge theory calculations for the
lightly doped strong coupling $t$-$J$ model by Lee and Wen \cite{Lee97}.
Signs of such behavior are also evident in a recent analysis of the
momentum distribution using a high temperature series by
Putikka {\em et al.}\cite{Putikka98x}.
Note models which include only n.n. hoppings
($t'=0$) are a special limit from the present
point of view.

The development of ISL near the saddle points 
is related to the gap formation of the high-$T_{\rm c}$ cuprates.
In the normal state of the underdoped cuprates, the
ARPES experiments \cite{Loeser96,Norman98}
show a single particle gap 
opening and a loss of quasiparticle weight 
in the vicinity of the saddle points below the
pseudo-gap temperature.
Tunneling experiment \cite{Renner98} also shows 
a quasi-particle gap formation above $T_{\rm c}$.
These results are quite similar to those we propose in Fig.~\ref{FigFS}b.
Systematics of the loss of quasiparticle weights in
electron- and hole-doped cuprates \cite{Kim98}
are also consistent with our results for 4-patch and 2-patch models.


The proposal that an ISL truncates the FS along the US
in the vicinity of the saddle points has very interesting consequences.
There will be a coupling to the open segments in the Cooper channel
through the scattering of electron pairs out of the ISL to the
open FS segments. This process is reminiscent of the coupling of
fermions to bosonic preformed pairs in the 
Geshkenbein-Ioffe-Larkin model \cite{Geshkenbein97}.
 They argued for an infinite mass for such
pairs to suppress their contribution to transport properties. Such
scattering processes will be an efficient mechanism for d-wave
pairing on the open FS segments.

We also see a close similarity to a phenomenological model
by Ioffe and Millis \cite{Ioffe98x},
to explain the anomalous transport properties  in the normal state.
They assumed the FS segments to have usual quasi-particle properties
without spin-charge separation
but the scattering rate  to vary strongly along the FS arcs.
They justified the model by comparison to 
ARPES and tunneling experiments \cite{Loeser96,Norman98,Renner98}.
In our case,
the scattering rate will vary strongly due to the strong Umklapp scattering
at the end of the FS arcs  where they meet the ISL region.
Lastly we refer the reader to the recent paper by
Balents, Fisher and Nayak\cite{Balents98x}
which introduces the concept of a Nodal liquid 
with properties similar to the ISL discussed above.

In conclusion we have shown that
when the Fermi surface approaches 
the saddle points, Umklapp scattering 
drives the system into a strong coupling fixed point which
 can cause a breakdown of Landau Fermi liquid state.
 The Fermi surfaces near the saddle points
$(\pi,0)$ and $(0,\pi)$ are truncated by the
formation of a pinned and insulating condensate, while
the zone diagonal regions around $(\pm\pi/2,\pm\pi/2)$ remain metallic.
 We have
given arguments that the spin properties are those of an insulating
spin liquid. This microscopic model has a lot in common with 
the results of ARPES experiments and some
recent phenomenological models so that we believe it can form the
basis for a theory of the cuprates. 

We wish to thank S. Haas, D. Khveshchenko, M. Sigrist, E. Trubowitz,
F.C. Zhang and R. Hlubina for stimulating conversations.
We acknowledge D. Poilblanc for discussion and
for pointing out the reference \cite{Lederer87}.
N.F. is supported by a Monbusho Grant for overseas research.



\begin{references}

\bibitem[*]{PermAddress} On leave from
Institute for Solid State Physics, Univ.~of Tokyo,
Minato-ku, Tokyo 106-8666, Japan. 


\bibitem{Furukawa98}
N. Furukawa and T. M. Rice,
J. Phys. Cond. Mat. {\bf 10}, L381 (1998) 

\bibitem{Schulz87}
H.J. Schulz, Europhys. Lett. {\bf 4}, 609 (1987).

\bibitem{Dzyaloshinskii87}
I. Dzyaloshinskii, Sov. Phys. JETP {\bf 66}, 848 (1987).

\bibitem{Lederer87}
P. Lederer, G. Montambaux and D. Poilblanc,
J. Physique {\bf 48}, 1613 (1987).

\bibitem{Dzyaloshinskii96}
I. Dzyaloshinskii, J. Phys. I France {\bf 6}, 119 (1996).



\bibitem{Feldman96} J. Feldman, M. Salmhofer and E. Trubowitz,
J. Stat. Phys. {\bf 84}, 1209 (1996).

\bibitem{Balents96} L. Balents and M.P.A. Fisher,
Phys. Rev. B{\bf 53}, 12133 (1996); H.-H. Lin, L. Balents and M.P.A. Fisher,
Phys. Rev. B{\bf 56}, 6569 (1997).

\bibitem{Rice97} \ T.M. Rice, S. Haas, M. Sigrist and F.C. Zhang,
Phys. Rev. B{\bf 56} 14655 (1997).

\bibitem{White98} \ S.R. White and D.J. Scalapino,
Phys. Rev. B{\bf 57}, 3031 (1998).

\bibitem{Lee97} P. A. Lee and X. G. Wen, Phys. Rev. Lett.
{\bf 78}, 4111 (1997).

\bibitem{Putikka98x} W. O. Putikka, M. U. Luchini and R. R. P. Singh,
preprint, cond-mat/9803141.



\bibitem{Loeser96} 
A.G. Loeser {\em et al.},
 Science {\bf 273}, 325 (1996).

\bibitem{Norman98}
M.R. Norman {\em et al.},
Nature {\bf 392}, 157 (1998).

\bibitem{Renner98} 
Ch.~Renner {\em et al.},
 Phys. Rev. Lett. {\bf 80}, 149 (1998).


\bibitem{Kim98}
C.~Kim {\em et al.}, Phys. Rev. Lett. {\bf 80}, 4245 (1998).

\bibitem{Geshkenbein97} V.G. Geshkenbein, L.B. Ioffe and A.I. Larkin, 
Phys. Rev. B{\bf 55}, 3173 (1997).

\bibitem{Ioffe98x} L.B. Ioffe and A.J. Millis, preprint, cond-mat/9801092.


\bibitem{Balents98x}
L. Balents, M. P. A. Fisher and  C. Nayak,
preprint, cond-mat/9803086.

\end{references}


\end{document}